\documentclass[pre,twocolumn,showpacs]{revtex4-1}
\usepackage{graphicx}
\usepackage{amsmath,amssymb}
\usepackage{color}
\usepackage{upgreek} 

\DeclareMathOperator{\erf}{erf}

\newcommand{\um}{\upmu\mathrm{m}}
\newcommand{\umpersec}{\um\,\mathrm{s}^{-1}}

\newcommand{\mV}{\mathrm{m\kern-.05em{}V}} 

\newcommand{\kB}{k_{\mathrm{B}}}
\newcommand{\kT}{\kB T}
\newcommand{\zetanondim}{{\overline{\zeta}}}
\newcommand{\myvec}[1]{\mathbf{#1}}
\newcommand{\Jvec}{\myvec{J}}
\newcommand{\Ivec}{\myvec{I}}
\newcommand{\Evec}{\myvec{E}}
\newcommand{\Uvec}{\myvec{U}}
\newcommand{\Rvec}{\myvec{R}}
\newcommand{\nvec}{\myvec{n}}
\newcommand{\rvec}{\myvec{r}}
\newcommand{\zerovec}{0}

\newcommand{\mob}{{\mathcal{M}}}

\newcommand{\rhoz}{\rho_z}

\newcommand{\leibnizd}{{\mathrm d}}
\newcommand{\dV}{\leibnizd V}
\newcommand{\dS}{\leibnizd S}
\newcommand{\dRdt}{\leibnizd\Rvec/\leibnizd t}

\newcommand{\half}{{\textstyle\frac{1}{2}}}

\newcommand{\lD}{\lambda_{\text{D}}}
\newcommand{\rhoD}{\rho_{\text{D}}}

\newcommand{\epsr}{\epsilon} 

\newcommand{\textint}{{\textstyle\int}}

\newcommand{\latin}[1]{{\itshape #1}}

\newcommand{\ie}{\latin{i.\,e.}}

\newcommand{\via}{\latin{via}}

\newcommand{\modulo}{\latin{modulo}}

\newcommand{\faradaic}{Faradaic} 
\newcommand{\ohmic}{Ohmic} 

\newcommand{\zetapotential}{zeta potential} 
\newcommand{\zetapotentials}{\zetapotential s}

\newcommand{\Eqref}[1]{Eq.~\eqref{#1}}
\newcommand{\Eqsref}[1]{Eqs.~\eqref{#1}}

\newcommand{\Figref}[1]{Fig.\@ \ref{#1}}

\newcommand{\figwidth}{0.95\linewidth} 

\begin{document}

\title{Non-\faradaic\ electric currents in the Nernst-Planck
  equations\\ and `action at a distance'
  diffusiophoresis in crossed salt gradients}

\author{Patrick B. Warren}

\email{patrick.warren@stfc.ac.uk}

\affiliation{STFC Hartree Centre, Scitech Daresbury,
  Warrington, WA4 4AD, UK}

\affiliation{Unilever R\&D Port Sunlight, Quarry Road East, Bebington,
  Wirral, CH63 3JW, UK.}

\date{December 11, 2019}

\begin{abstract}
  In the Nernst-Planck equations in two or more dimensions, a
  non-\faradaic\ electric current can arise as a consequence of
  connecting patches with different liquid junction potentials.
  Whereas this current vanishes for binary electrolytes or
  one-dimensional problems, it is in general non-vanishing for example
  in crossed salt gradients.  For a suspended colloidal particle,
  electrophoresis in the corresponding electrostatic potential
  gradient is generally vectorially misaligned with chemiphoresis in
  the concentration gradients, and diffusiophoresis
  (\via\ electrophoresis) can occur in regions where there are no
  local concentration gradients (`action at a distance').  These
  phenomena may provide new opportunities to manipulate and sort
  particles, in microfluidic devices for example.
\end{abstract}

\maketitle

The growing realisation that diffusiophoresis is a potent and
ubiquitous non-equilibrium transport mechanism for micron-sized
colloidal particles has led to a recent surge of interest in the
phenomenon \cite{ACY+08, PCY+12, RSP13, FMH+14, SNA+16, BWA+16, Keh16,
  VGG+16, SSW+17, SAW+17}.  For example, diffusiophoresis is
effective at injecting or ousting particles from dead-end channels
\cite{KCR+15, SUS+16}, has been identified as a hitherto unsuspected
pore-scale particulate soil removal process in laundry detergency \cite{SWS18},
implicated as a general non-motor transport mechanism in cells \cite{Sea19},
and can be used to manipulate and sort particles by size and charge
\cite{SUS+16, SAF+17}.  The biggest effects arise in electrolyte
solutions, where chemiphoresis in concentration gradients combines
with electrophoresis in the diffusion potential to drive particles at
speeds of 1--10\,$\umpersec$ \cite{And89}, propelling them over large
distances in time scales of minutes.  An additional peculiarity in
binary electrolytes is that the speed is logarithmically dependent on
the concentration, leading to persistent effects such as osmotic
trapping \cite{PCY+12}, and long-lived particle removal \cite{SWS18}.

To my knowledge, the existing phenomena that have been discussed in
the above context pertain to binary electrolytes or assume
one-dimensional gradients \cite{BP14, CV14, SNA+16, GSI+19, GRS19}.
In this article, I argue that a still further enriched phenomenology
arises in multicomponent electrolytes when concentration gradients are
superimposed in different directions (`crossed' salt gradients).  In
part this is because chemiphoresis decouples partially from
electrophoresis, but additionally it is because a non-vanishing
electric current arises even in the absence of \faradaic\ reactions,
when patches with different liquid junction potentials are
connected by the intervening electrolyte solution.  In itself
this is surely a fascinating phenomenon, but importantly for
diffusiophoresis, the presence of electric fields in bulk regions
where there are no local concentration gradients implies that
particles should move in those regions, as a kind of diffusiophoretic
`action at a distance'.  Since it seems quite easy to engineer crossed
gradients either in microfluidics devices or with suitably chosen
`soluto-inertial beacons' as sources and sinks \cite{BWA+16, BS19}, these
observations provide novel opportunities for particle manipulation and
sorting.

Let me start with the Nernst-Planck equations which govern ion
transport in these problems \cite{Lev62, NT04},
\begin{equation}
\frac{\partial \rho_i}{\partial t}+\nabla\cdot\Jvec_i=0\,,\quad
\Jvec_i=-D_i(\nabla \rho_i + \rho_i z_i \nabla\varphi)\,.\label{eq:np}
\end{equation}
In these, $\rho_i$ is the density of the $i$-th ionic species, $D_i$
is the corresponding diffusion coefficient, $z_i$ the charge on the
ion in units of $e$, where $e$ is the unit of elementary charge, and
$\varphi=e\phi/\kT$ is a dimensionless electrostatic potential wherein
$\kB T$ is the unit of thermal energy and $\phi$ is the actual
electrostatic potential.  \Eqsref{eq:np} combine mass conservation
laws for the individual ion densities with expressions for the fluxes
driven by diffusion and drift in the electric field.  For simplicity I
omit advection terms although these are certainly relevant in
microfluidics devices, and may additially arise if bulk flows are
driven by diffusio-osmotic effects \cite{SUS+16}.

The Nernst-Planck equations must be augmented by a closure for the
electrostatic potential.  At a fundamental level this is the Poisson
equation, $\epsr\nabla^2\phi=-e\rhoz$, where $\rhoz=\sum_i z_i \rho_i$
is the space charge (in units of $e$) and $\epsr$ is the permittivity
(assumed constant) of the supporting medium. The combined set are then
known as the Poisson-Nernst-Planck (PNP) equations.  Introducing the
Debye length $\lD = (\epsr\kT/e^2\!\rhoD)^{1/2}$, where $\rhoD =
\sum_iz_i^2\rho_i$, allows the Poisson equation to be written as
$\lD^{2}\,\nabla^2\varphi=-{\rhoz}/{\rhoD}$.  This makes it clear that
if the problem size $L\gg\lD$, the bare electrostatics problem is
\emph{singular} \cite{Haf65, Hic70, Jac74, AMP87, BTA04, JB18}, in the
sense that there is an `outer' domain on the length scale $O(L)$ in
which $\rhoz\approx0$ (local charge neutrality), asymptotically
matched to `inner' solutions on a length scale $O(\lD)$ (\ie\ electric
double layers or EDLs), whenever the boundary conditions would
otherwise over-determine $\varphi$ in the outer domain
\cite{edl-note}.

Crucially, local charge neutrality does \emph{not} necessarily imply a
vanishing electric current $\Ivec=\sum_i\,z_i\Jvec_i$ (in units of
$e$) in the outer domain.  Rather, by summing the mass conservation
laws in \Eqsref{eq:np} one can only conclude that the current should
be \emph{solenoidal} ($\nabla\cdot\Ivec=0$).  In fact, even for pure
diffusion problems without \faradaic\ reactions \cite{dhukin-note}, a
non-vanishing current ($\Ivec\ne\zerovec$) is not only possible but may be
\emph{mandatory}.  To see this, insert the fluxes $\Jvec_i$ from
\Eqsref{eq:np} into the definition of $\Ivec$ to obtain
\begin{equation}
  \Ivec=-\nabla g + \sigma\Evec\,.\label{eq:ieq}
\end{equation}
This decomposes $\Ivec$ into the sum of a diffusion current, and a
conduction current obeying Ohm's law \cite{NT04}. In this
$g={\textstyle\sum_i}\,z_iD_i\rho_i$ is a weighted sum of ion
densities, $\sigma = {\textstyle\sum_i}\,z_i^2 D_i \rho_i$ is the
conductivity, and $\Evec=-\nabla\varphi$ is the electric field (the
latter two are in semi-reduced units).  Proceeding from
\Eqref{eq:ieq}, if $\Ivec=\zerovec$ then it is easy to show
$\nabla\times\Evec=\sigma^{-2}\,{\nabla\sigma\times\nabla g}$.  But
there is no particular reason why the cross product on the right hand
side should vanish, even though $\nabla\times\Evec=\zerovec$ because
$\Evec=-\nabla\varphi$.  Thus we are forced to conclude that in
general $\Ivec\ne\zerovec$.  As another way to see this, by taking the
curl of \Eqref{eq:ieq} one can eliminate the electrostatic potential
to find
\begin{equation}
  \sigma\,\nabla\times\Ivec=
  \nabla\sigma\times(\Ivec+\nabla g)\,.\label{eq:icurl}
\end{equation}
This is an inhomogeneous partial differential equation for $\Ivec$,
and again supports the notion that $\Ivec\ne\zerovec$ is driven by
crossed gradients in the form $\nabla\sigma\times\nabla g\ne0$.

\begin{figure}
\centering
  \includegraphics[clip=true,width=\figwidth]{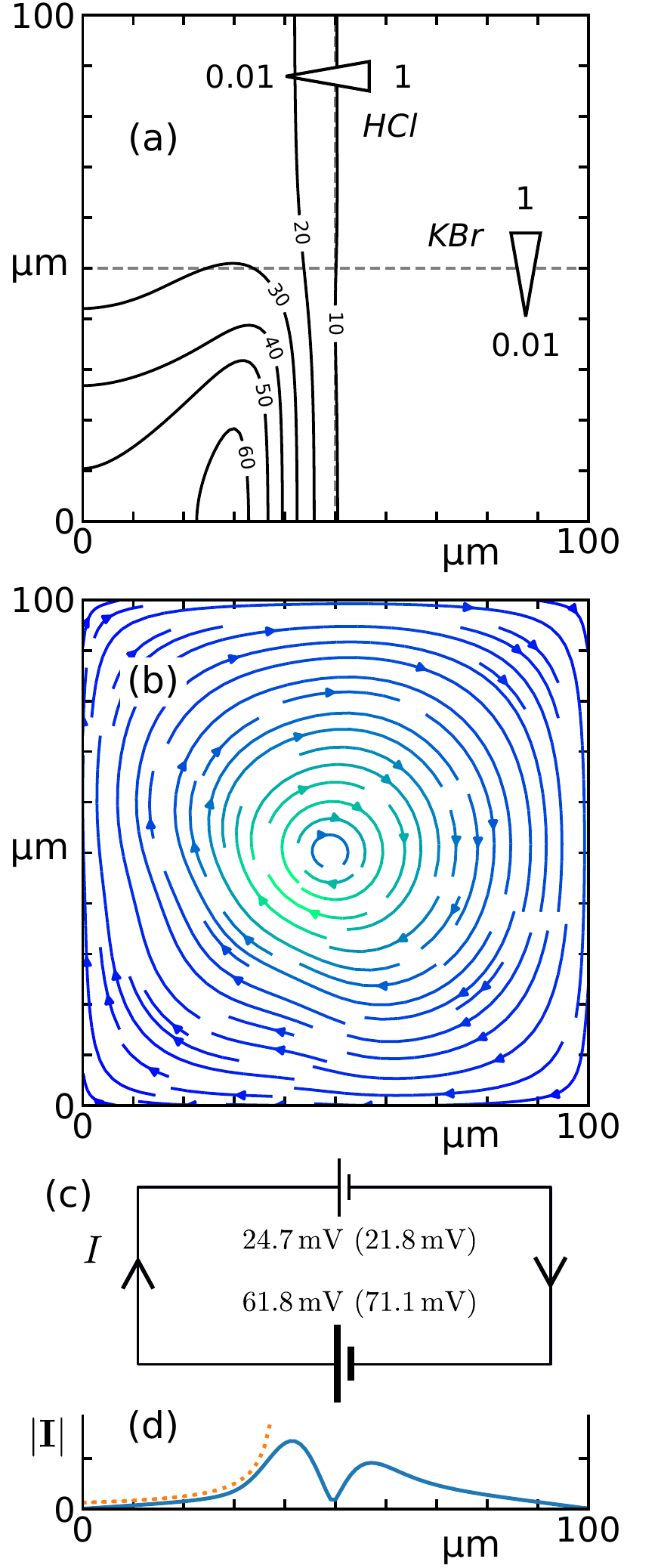}
  \caption{Crossed salt gradients at $t=0$, between HCl (horizontal)
    and KBr (vertical) \cite{shape-note}: (a) electrostatic potential
    (equipotential lines labelled in $\mV$) and space charge (colored
    background); (b) circulating electric current (colored by
    magnitude); (c) equivalent circuit with HCl gradients labelled by
    liquid junction potentials (in brackets are the `open circuit'
    values calculated from $\Delta\varphi =
    -\beta\ln(\sigma_2/\sigma_1)$ where $\sigma_2/\sigma_1$ is the
    ratio of conductivities across the junction \cite{binary-note});
    (d) magnitude of current along the $x=y$ diagonal (the dotted line
    is $1/r\sim1/|x-x_0|$ where $x_0=40\,\um$).\label{fig:t0combi}}
\end{figure}

Of course there are many examples where $\Ivec$ does vanish.  One
such case is where the gradients are one-dimensional so that $\varphi$
can be found by quadrature \cite{GSI+19}.  Another important case
is that of a \emph{binary} electrolyte \cite{Lev62, NT04} for which
$\varphi=-\beta\ln\rho_s$ (the diffusion or \emph{liquid junction
  potential}).  Here $\beta={(D_1-D_2)}{(q_1D_1+q_2D_2)}$ is a
normalised diffusivity contrast, and I suppose that $z_1>0$ and
$z_2<0$, set $q_i=|z_i|$, and use $\rho_s = \rho_1/q_2 = \rho_2/q_1$
for the overall electrolyte concentration \cite{binary-note}.

The simplest situation where an electric current \emph{does} arise is
where there are three ion species, with crossed gradients.  To explore
this, suppose there are two cations with a common anion.  Let the
respective ion densities be $\rho_1$, $\rho_2$ and $\rho_0$, with
corresponding diffusivities $D_1$, $D_2$ and $D_0$, and let the ions
be univalent ($|z_i|=1$).  For local charge neutrality we have
$\rho_0=\rho_1+\rho_2$.  Then $\nabla\sigma\times\nabla
g=2D_0(D_2-D_1) \nabla \rho_1\times\nabla \rho_2$.  This suggests that
the appearance of an electric current requires crossed gradients
\emph{and} contrasting cation diffusivities ($D_1\ne D_2$), but no
particular requirement is placed on the anion diffusivity.  Thus one
of the gradients can be in a supporting electrolyte
(\ie\ $\beta\approx0$), as in the example below.

To summarise the mathematical problem thus far, given $g$ and $\sigma$
and supposing that $\Ivec\cdot\nvec$ is specified on the boundaries of
the domain of interest, we must find the current distribution that
satisfies \Eqref{eq:ieq} with $\nabla\cdot\Ivec=0$ and
$\Evec=-\nabla\varphi$.  To prove solutions do exist, and are unique,
we can note that this combination implies \cite{NT04, RB10}
\begin{equation}
  \nabla\cdot(\sigma\nabla\varphi) + \nabla^2\!g=0\,.\label{eq:ihp}
\end{equation}
This is an inhomogeneous Poisson equation for $\varphi$ with the
equivalent of a spatially-varying dielectric permittivity.  Existence
and uniqueness of $\varphi$ (up to an additive constant) then follows
by analogy with standard electrostatics \cite{Cou61}. A direct proof
is also given in Appendix \ref{app:uniq}.
\Eqref{eq:ihp} is non-singular and amenable to solution by standard
numerical methods \cite{PTV+07}, and replaces the original
electrostatic Poisson equation in closing the Nernst-Planck equations.
Additionally, I show in Appendix \ref{app:var} that the variational
principle equivalent to \Eqref{eq:ihp} corresponds to minimising the
total \ohmic\ heating $\int \Ivec^2/2\sigma\,\dV$ \modulo\ a surface
term.  Recalling that the problem is athermal, this can be interpreted
as a proxy minimum entropy production principle.  The connection to
the true entropy production in the underlying PNP equations is left
for future work.

\begin{figure}[t]
\centering
  \includegraphics[clip=true,width=\figwidth]{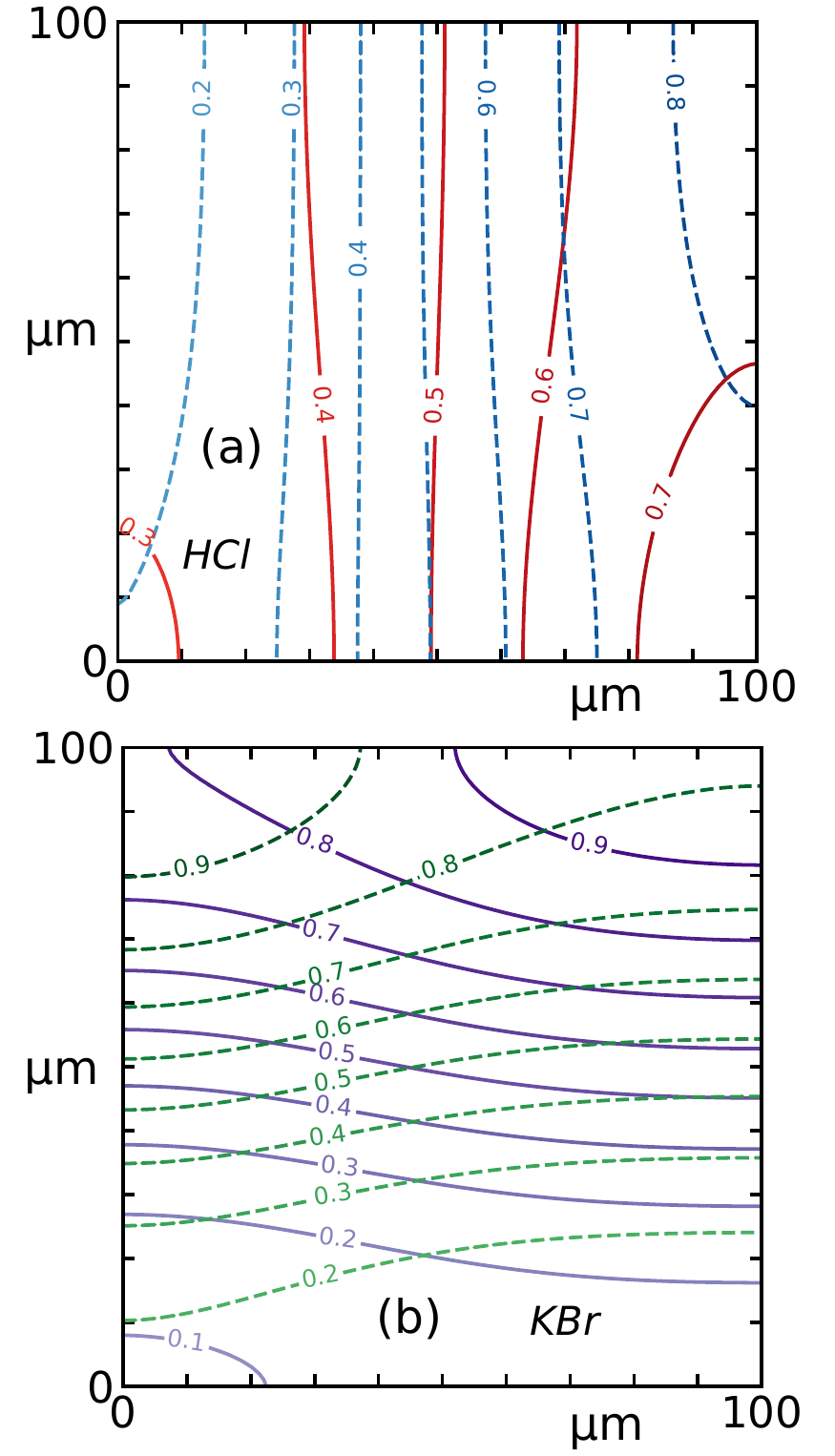}
  \caption{Ion density profiles at $t=250\,\mathrm{ms}$, for (a) HCl
    and (b) KBr.  Contours are shown as solid and dashed lines for
    cations and anions respectively, labelled by concentration using the
    same scale as \Figref{fig:t0combi}a.\label{fig:t250}}
\end{figure}

Let me turn now to a specific numerical example which demonstrates the
principles by which a non-vanishing electric current arises.  For
concreteness I consider an enclosed square domain of side 100\,$\um$,
initialised at $t=0$ with a 100-fold gradient in an electrolyte with a
large diffusivity contrast (HCl, $\beta=0.64$), crossed with a
100-fold gradient in a supporting electrolyte (KBr, $\beta=-0.01$)
\cite{diffs-note}.  The concentration gradients are initially
localised to the mid-planes, with widths $10\,\um$ \cite{shape-note},
so that the square domain is divided into four quadrants as shown in
\Figref{fig:t0combi}a.  The actual concentration units need not be
specified since the overall units of concentration can be factored out
of the Nernst-Planck equations.  For this demonstration I choose a
problem with four rather than three ions, since this maintains the
distinction between the two electrolytes.

I solve \Eqref{eq:ihp} in this square domain, with $\Ivec\cdot\nvec=0$
on the boundaries.  For details see Appendix \ref{app:num}.
\Figref{fig:t0combi}a shows that there is a significant liquid
junction potential ($\Delta\phi\approx 62\,\mV$) between the two lower
quadrants, corresponding approximately to the expected value for HCl
treated as a binary electrolyte.  The junction potential between the
upper two quadrants is much weaker though ($\Delta\phi\approx
25\,\mV$), as might be expected for HCl in the presence of a
supporting electrolyte \cite{binary-note}.  It is essentially this
difference that drives the circulating electric current
(\Figref{fig:t0combi}b).  By joining the upper and lower halves, it is
as if we have \emph{short-circuited} the two liquid junctions, as
sketched in \Figref{fig:t0combi}c.  The resulting current is
distributed throughout the square domain, as befits the minimum
\ohmic\ heating principle.  Crucially, in the lower-left quadrant
where the conductivity is small, this generates a significant electric
field \emph{throughout} this region as indicated by the equipotential
lines in \Figref{fig:t0combi}a.  Also shown in \Figref{fig:t0combi}a
is the space charge from $\rhoz=-(\epsr\kT/e^2)\,\nabla^2\varphi$.
Note that $|\rhoz|\alt 10^{-8}\,\mathrm{M}$ so that local charge
neutrality should normally be a very good approximation
\cite{rhoz-note}.  Finally, \Eqref{eq:icurl} implies $\Ivec$ should be
\emph{irrotational} as well as solenoidal, in regions where the
gradients vanish.  This explains why approximately $|\Ivec|\sim1/r$ in
the lower left quadrant (\Figref{fig:t0combi}d), and why the
equipotential lines are approximately radial in this quadrant
(\Figref{fig:t0combi}a).

\begin{figure}[t]
\centering
  \includegraphics[clip=true,width=\figwidth]{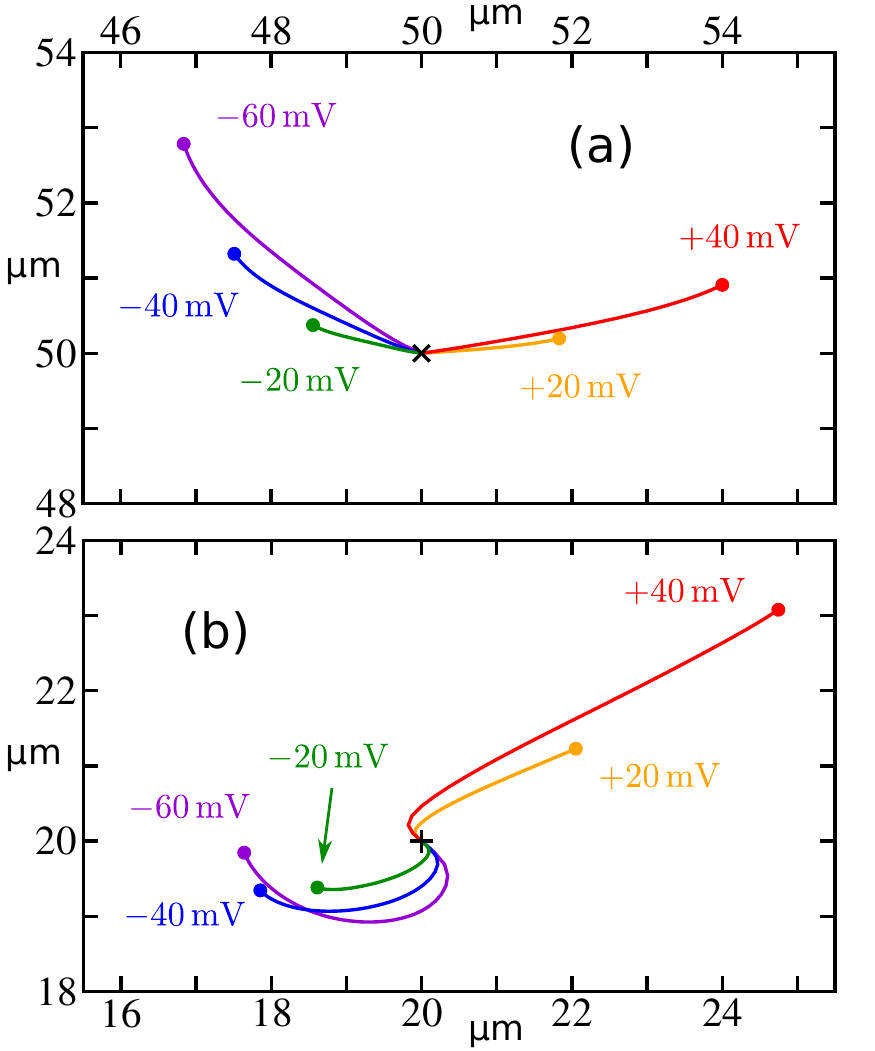}
  \caption{Diffusiophoresis: trajectories of colloid particles with
    different \zetapotential{}s (labels) starting from (a) the
    centre of the crossed gradients, and (b) a location in the lower
    left quadrant.  Starting points are marked by a cross, and
    positions at $t=500\,\mathrm{ms}$ by filled
    circles.\label{fig:traj}}
\end{figure}

As time progresses, the gradients in this confined system dissipate by
coupled diffusion.  To track the evolving concentration fields, I
solve the Nernst-Planck equations with boundary conditions
$\Jvec_i\cdot\nvec=0$, computing the electrostatic potential from
\Eqref{eq:ihp} at each step \cite{timescale-note}.  \Figref{fig:t250}
shows the situation after 250\,ms.  In the upper half space the more
mobile $\mathrm{H}^+$ has spread out much further than the less mobile
$\mathrm{Cl}^-$ (\Figref{fig:t250}a), since with the high
concentration of KBr in this region the ion densities become
decoupled.  Additionally the circulating current corresponds to
cations moving clockwise and anions moving anticlockwise, which
distorts the ion density profiles, as seen for
$\mathrm{K}^+$ and $\mathrm{Br}^-$ (\Figref{fig:t250}b).

What are the implications for diffusiophoresis of a suspended
colloidal particle?  Obviously, this depends on where the particle is
located as well as its \zetapotential.  Here I predict trajectories by
integrating $\dRdt=\Uvec$, where the diffusiophoretic drift
velocity is \cite{And89, SNA+16, GRS19, uni-note}
\begin{equation}
  \Uvec=\frac{\epsr}{\eta}\,\Bigl(\frac{\kT}{e}\Bigr)^2
  \Bigl[4\ln\cosh\Bigl(\frac{\zetanondim}{4}\Bigr)\,\nabla\ln\rho
  -\zetanondim\,\nabla\varphi\Bigr]\label{eq:dp}
\end{equation}
(see also Appendix \ref{app:dp}).  In this $\rho=\sum_i\rho_i$ is the
total ion density, $\eta$ is the viscosity of the
medium, and $\zetanondim=e\zeta/\kT$ is the
non-dimensionalised \zetapotential.  The two terms in \Eqref{eq:dp}
correspond respectively to chemiphoresis in the overall concentration
gradient, and electrophoresis in the electrostatic potential gradient.

Sample trajectories are shown in \Figref{fig:traj}.  In the lower-left
quadrant (\Figref{fig:traj}b) the electric field corresponding to the
gradient in $\phi$ drives diffusiophoresis \via\ electrophoresis even
though there are initially \emph{no} local concentration gradients.  I
term this unusual phenomenon diffusiophoretic `action at a distance'.
Since the electric field \emph{also} drives the electric current, in
this quadrant $\Uvec$ is initally parallel to $\Ivec$; this explains
the initial coincidence of the trajectories.  In contrast, for a
particle which finds itself in the middle of the crossed salt
gradients (\Figref{fig:traj}a), electrophoresis and chemiphoresis are
vectorially misaligned even initially, so that particles with
different \zetapotentials\ are propelled along diverging trajectories
even if they have the same sign of charge.

The design of devices which exploit these striking effects is a
clearly a promising avenue for future work.  I note that in this
situation one loses the logarithmic sensitivity exhibited in binary
electrolytes \cite{PCY+12, SWS18, logsens-note}, so that the distance
over which particles move is limited by the relaxation time for the
ion densities.  This can be alleviated by using soluto-inertial
beacons \cite{BWA+16, BS19}, or microfluidic devices in which
long-lived gradients can be established \cite{ACY+08, SNA+16, SSW+17}.

To summarise, a rich phenomenology arises in the Nernst-Planck
equations when considering multicomponent electrolytes in more than
one dimension.  In particular, circulating (solenoidal) electric
currents appear when patches with different liquid junction potentials
are connected by the intervening electrolyte solution.  The electric
fields associated with these currents can drive `action at a distance'
diffusiophoresis of suspended colloidal particles, even in the absence
of local concentration gradients.  This is a definitive prediction of
the Nernst-Planck equations, combined with the current understanding of
diffusiophoresis of charged colloidal particles, and it
would be fascinating to put to an experimental test.

\acknowledgments

I thank Sangwoo Shin and Howard A.\ Stone for a critical reading of
the draft manuscript.

\appendix

\section{Uniqueness}\label{app:uniq}
Here I provide a direct proof of uniqueness of $\varphi$ in
\Eqref{eq:ihp}.  Suppose there are two solution pairs
$(\Ivec_1,\varphi_1)$ and $(\Ivec_2,\varphi_2)$, such that
$\Ivec_1\cdot\nvec=\Ivec_2\cdot\nvec$ on some domain boundary with
vector normal $\nvec$.  Subtracting the corresponding versions of
\Eqref{eq:ieq} yields a homogeneous problem in which the difference
solution, with $\Ivec=\Ivec_2-\Ivec_1$ and
$\varphi=\varphi_2-\varphi_1$, satisfies Ohm's law $\Ivec=\sigma\Evec$
where $\Ivec\cdot\nvec=0$ on the domain boundary, $\nabla\cdot\Ivec=0$
in the interior, and $\Evec=-\nabla\varphi$.  Now consider
\begin{equation}
  \nabla\cdot(\varphi\Ivec)=\varphi\,(\nabla\cdot\Ivec)
  +\Ivec\cdot\nabla\varphi\,.\label{eq:mid}
\end{equation}
The first term on the right hand side vanishes as a consequence of the
solenoidal nature of $\Ivec$, and the second term simplifies to
$\Ivec\cdot\nabla\varphi=-\sigma\Evec^2$.  Integrate \Eqref{eq:mid}
over the domain of interest and use the divergence theorem to get
\begin{equation}
  \textint \nabla\cdot(\varphi\Ivec)\,\dV
  =\textint \varphi\Ivec\cdot\nvec\,\dS=0
\end{equation}
(because $\Ivec\cdot\nvec=0$ on the boundary).  We conclude that
\begin{equation}
  \textint\sigma\Evec^2\,\dV=0\,.
\end{equation}
But $\sigma>0$ and $\Evec^2\ge0$, so this implies $\Evec=0$
everywhere, and hence $\Ivec=0$ and $\varphi=\mathrm{constant}$.  This
is the desired result.  It means that the solution pairs
$(\Ivec_1,\varphi_1)$ and $(\Ivec_2,\varphi_2)$ in the original
inhomogeneous problem can at most differ by a constant in $\varphi$.

\section{Variational principle}\label{app:var}
An inhomogeneous Poisson equation such as that given in 
\Eqref{eq:ihp} has an equivalent variational principle.  In the
present case it is
\begin{equation}
  \frac{\delta }{\delta\varphi(\rvec)}\Bigl[
  \textint (\half\sigma|\nabla\varphi|^2-\varphi\nabla^2g)\,\dV\Bigr]=0\,.
\end{equation}
Making use of the vector calculus identity
\begin{equation}
\nabla\cdot(\varphi\nabla g)=\nabla\varphi\cdot\nabla g
+\varphi\nabla^2 g\,,
\end{equation}
the integrand in the above can be reformulated to
\begin{equation}
  \begin{split}
    &\half{\sigma|\nabla\varphi|^2}-\varphi\nabla^2g\\[3pt]
    &\qquad{}=\frac{|\sigma\nabla\varphi+\nabla g|^2}{2\sigma}
    -\frac{|\nabla g|^2}{2\sigma}-\nabla\cdot(\varphi\nabla g)\,.
    \end{split}
\end{equation}
The second term on the right hand side is constant, given $g$ and
$\sigma$, and can be discarded.  The third term can be replaced by
a surface integral.  Rewriting in terms of the currents, the
variational principle can be rebranded as
\begin{equation}
  \frac{\delta}{\delta\varphi(\rvec)}
  \Bigl[\textint ({\Ivec^2}/{2\sigma})\,\dV
    -\textint\!\varphi\,\nabla g\cdot\nvec\,\dS\Bigr]=0\,.
\end{equation}
Thus the electrostatic potential in \Eqref{eq:ihp} is such as to
minimise the \ohmic\ heating (\ie\ defined using the total
current), \modulo\ a surface term.

\begin{figure}
\begin{center}
  \includegraphics[clip=true,width=\figwidth]{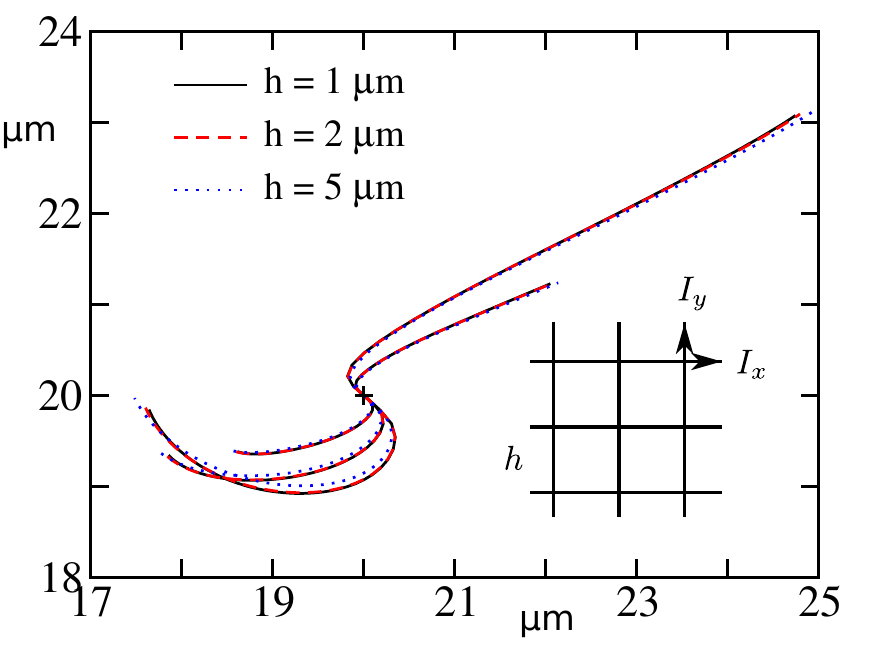}
\end{center}
\vskip-18pt
\caption{Dependence of trajectories in \Figref{fig:traj}b on grid
  spacing.\label{fig:numerics}}
\end{figure}

\section{Numerical scheme}\label{app:num}
To solve the inhomogeneous Poisson equation, \Eqref{eq:ihp}, I
discretise the problem domain into a square grid of spacing $h$
(\Figref{fig:numerics} inset).  The potential $\varphi$ and the ion
densities $\rho_i$ are defined on the nodes of the grid, and the
fluxes $\Jvec_i$ and current $\Ivec$ are defined on the edges joining
the nodes (shown in \Figref{fig:numerics} inset for $I_x$ and $I_y$
components).  With these definitions, $\nabla\cdot\Ivec=0$ becomes the
constraint that the sum of the currents entering each node should
vanish.  The number of constraints then matches the number unknowns
(values of $\phi$ on the nodes) and the problem is linear, so in
principle can be solved by any (sparse) linear algebra method.  In
practice I use a straightforward Gauss-Seidel iterative scheme that
requires minimal bookkeeping, with the convergence criterion being
that the relative change in $\varphi$ in subsequent iteractions falls
to less than $10^{-14}$.  For boundary conditions I set the fluxes to
zero on the exterior edges.  Note that the actual space charge is not
represented as such in the calculation, and deviations from
$\nabla\cdot\Ivec=0$ are numerical errors.

To solve the time-dependent Nernst-Planck equations, I use a standard
forward-time centered-space (FTCS) scheme \cite{PTV+07} based on the
above grid decomposition, with a time step $\delta t =
0.025\,\mathrm{ms}\times (h/\um)^2$, which comfortably satisfies the
usual Courant-Friedrichs-Lewy condition since the maximum diffusion
coefficient is $D=9.31\,\um^2\,\mathrm{ms}^{-1}$ (for $\mathrm{H}^+$)
so $D\,\delta t/h^2<0.5$.

To compute the trajectories of particles undergoing diffusiophoresis I
integrate the kinematic equations in \Eqref{eq:dp} using a simple
first order forward Euler scheme with $\Delta t=5\,\mathrm{ms}$ (a
multiple of $\delta t$), and bivariate spline interpolation (on the
same grid as above) to calculate off-lattice approximations to
$\nabla\ln\rho$ and $\nabla\varphi$.

\Figref{fig:numerics} shows that the computed trajectories depend very
little on the underlying grid spacing and consequent choices for time
step.  I only show the $h$-dependence for trajectories of particles
starting in the lower left quadrant; the trajectories of particles
which start in the centre of the crossed gradients show even smaller
$h$-dependence.  All calculations reported in the main text are for
$h=1\,\um$ (100$^2$ grid).

\section{Diffusiophoretic drift coefficients}\label{app:dp}
Diffusiophoresis in multicomponent electrolytes has been considered by
several groups recently \cite{CV14, SNA+16, GRS19}.  Assuming a thin EDL, it
is convenient to start with a general expression for the
diffusiophoretic drift of a suspended colloidal particle arising from
bulk chemical potential gradients,
\begin{equation}
  \Uvec={\textstyle\sum_i} \mob_i\nabla\mu_i\,.\label{supp:eq:udp}
\end{equation}
Restricting the analysis to the tractable but practically relevant
case of monovalent electrolytes, the mobilities are
$\mob_i=(\rho_i/\rho)\times\mob_\pm$ where \cite{SNA+16}
\begin{equation}
  \mob_\pm=\frac{\epsr\kT}{\eta e^2}\Bigr[
    4\ln\cosh\frac{e\zeta}{4\kT}\mp\frac{e\zeta}{\kT}\Bigr]
  \label{supp:eq:mob}
\end{equation}
according to the sign of the ion ($z_i=\pm1$).  Here
$\rho=\sum_i\rho_i$ as in the main text, $\eta$ is viscosity, and
$\zeta$ is the particle \zetapotential.  This formalism extends to
include electrophoresis if one employs the \emph{electrochemical}
potentials,
\begin{equation}
  \mu_i=\kT\ln\rho_i+ez_i\phi\,.
\end{equation}
Combining \Eqsref{supp:eq:udp} and \eqref{supp:eq:mob}, cross terms
cancel since $\sum_iz_i\rho_i=0$, yielding \Eqref{eq:dp} used in the
main text. Note that the second term in \Eqref{eq:dp} simplifies to
the well-known Helmholtz-Smoluchowski result
$-(\epsr\zeta/\eta)\nabla\phi$ \cite{And89}.


\begin{thebibliography}{41}
\bibitem [{edl()}]{edl-note}%
  \BibitemOpen
  \href@noop {} {}\bibinfo {note} {The value of $\varphi $ where the outer
  solution meets the EDLs cannot be pre-determined, and indeed it is the
  mismatch between this and the true wall boundary condition that gives rise to
  an EDL in the first place.}\BibitemShut {Stop}%
\bibitem [{dhu()}]{dhukin-note}%
  \BibitemOpen
  \href@noop {} {}\bibinfo {note} {Technically I also assume small Dhukin
  number so that surface conduction in the EDLs can be neglected
  \cite{And89}.}\BibitemShut {Stop}%
\bibitem [{sha()}]{shape-note}%
  \BibitemOpen
  \href@noop {} {}\bibinfo {note} {The specific functional form is
  $c=c_0+(c_1-c_0)\erf(x/w)$ where $c_0=0.01$ and $c_1=1$ are the limiting
  concentrations in arbitrary units, $x$ is the distance from the mid-plane,
  and $w=10\,\um$ is the width.}\BibitemShut {Stop}%
\bibitem [{rho()}]{rhoz-note}%
  \BibitemOpen
  \href@noop {} {}\bibinfo {note} {Interestingly the choice of units for the
  ion densities does not affect the calculation of the space charge
  $\rhoz$.}\BibitemShut {Stop}%
\bibitem [{tim()}]{timescale-note}%
  \BibitemOpen
  \href@noop {} {}\bibinfo {note} {The ion densities relax on a time scale
  $O(L^2/D)\sim10\,\mathrm{s}$ since $L=100\,\um$ and $D \sim 10^{-9}\,
  \mathrm{m}^2\, \mathrm{s}^{-1}$. On the other hand, the time scale for the
  electrostatic potential $\varphi$ and electric current $\Ivec$ to relax is
  $O(\lD^2/D)$ \cite{Hic70, Jac74} (as an $RC$ circuit, the relevant
  capacitance is $O(\lD/L)$ smaller than in the EDL charging problem
  \cite{BTA04, JB18}). Then $L\gg\lD$ implies $\varphi$ and $\Ivec$ are slaved
  to the ion densities. Further investigation of this aspect is left to future
  work.}\BibitemShut {Stop}%
\bibitem [{uni()}]{uni-note}%
  \BibitemOpen
  \href@noop {} {}\bibinfo {note} {\Eqref{eq:dp} assumes the ions are univalent
  \cite{GRS19}.}\BibitemShut {Stop}%
\end{thebibliography}
%

\end{document}